\documentclass[aps,prl,preprint,nopacs,superscriptaddress]{revtex4}

\usepackage{graphicx}
\usepackage{verbatim}
\usepackage{mathrsfs}
\usepackage{amsmath,amsfonts,amssymb}
\usepackage{graphicx}

\def\3{2.8in}    
\def\2{2.5in}
\def\4{3.0in}

\def \beq {\begin{equation}}
\def \eeq {\end{equation}}
\begin{document}

\title{A strongly robust type-II Weyl fermion semimetal state in Ta$_{3}$S$_{2}$}

\author{Guoqing Chang$^*$}\affiliation{Centre for Advanced 2D Materials and Graphene Research Centre National University of Singapore, 6 Science Drive 2, Singapore 117546}\affiliation{Department of Physics, National University of Singapore, 2 Science Drive 3, Singapore 117542}
\author{Su-Yang Xu\footnote{These authors contributed equally to this work.}$^{\dag}$}\affiliation {Laboratory for Topological Quantum Matter and Spectroscopy (B7), Department of Physics, Princeton University, Princeton, New Jersey 08544, USA}
\author{Daniel S. Sanchez}\affiliation {Laboratory for Topological Quantum Matter and Spectroscopy (B7), Department of Physics, Princeton University, Princeton, New Jersey 08544, USA}

\author{Shin-Ming Huang}
\affiliation{Centre for Advanced 2D Materials and Graphene Research Centre National University of Singapore, 6 Science Drive 2, Singapore 117546}
\affiliation{Department of Physics, National University of Singapore, 2 Science Drive 3, Singapore 117542}

\author{Chi-Cheng Lee}
\affiliation{Centre for Advanced 2D Materials and Graphene Research Centre National University of Singapore, 6 Science Drive 2, Singapore 117546}
\affiliation{Department of Physics, National University of Singapore, 2 Science Drive 3, Singapore 117542}
\author{Tay-Rong Chang}\affiliation{Department of Physics, National Tsing Hua University, Hsinchu 30013, Taiwan}
\affiliation {Laboratory for Topological Quantum Matter and Spectroscopy (B7), Department of Physics, Princeton University, Princeton, New Jersey 08544, USA}
\author{Guang Bian}\affiliation{Laboratory for Topological Quantum Matter and Spectroscopy (B7), Department of Physics, Princeton University, Princeton, New Jersey 08544, USA}

\author{Hao Zheng}\affiliation {Laboratory for Topological Quantum Matter and Spectroscopy (B7), Department of Physics, Princeton University, Princeton, New Jersey 08544, USA}

\author{Ilya Belopolski}\affiliation {Laboratory for Topological Quantum Matter and Spectroscopy (B7), Department of Physics, Princeton University, Princeton, New Jersey 08544, USA}

\author{Nasser Alidoust}\affiliation {Laboratory for Topological Quantum Matter and Spectroscopy (B7), Department of Physics, Princeton University, Princeton, New Jersey 08544, USA}


\author{Horng-Tay Jeng}\affiliation{Department of Physics, National Tsing Hua University, Hsinchu 30013, Taiwan}\affiliation{Institute of Physics, Academia Sinica, Taipei 11529, Taiwan}

\author{Arun Bansil}\affiliation{Department of Physics, Northeastern University, Boston, Massachusetts 02115, USA}

\author{Hsin Lin$^{\dag}$}
\affiliation{Centre for Advanced 2D Materials and Graphene Research Centre National University of Singapore, 6 Science Drive 2, Singapore 117546}
\affiliation{Department of Physics, National University of Singapore, 2 Science Drive 3, Singapore 117542}

\author{M. Zahid Hasan\footnote{Corresponding authors: suyangxu@princeton.edu, nilnish@gmail.com, and mzhasan@princeton.edu}}\affiliation {Laboratory for Topological Quantum Matter and Spectroscopy (B7), Department of Physics, Princeton University, Princeton, New Jersey 08544, USA}

\date{\today}

\begin{abstract}
Weyl semimetals are of great interest because they provide the first realization of the Weyl fermion, exhibit exotic quantum anomalies, and host Fermi arc surface states. The separation between Weyl nodes of opposite chirality gives a measure of the robustness of the Weyl semimetal state. In order to exploit the novel phenomena that arise from Weyl fermions in applications, it is crucially important to find robust Weyl semimetals with well-separated Weyl nodes. In this paper, we propose a methodology to design such materials. Using this methodology as a guideline, we search among the material parameter space and identify by far the most robust Weyl semimetal in the single crystalline compound tantalum sulfide Ta$_3$S$_2$ with new and novel properties beyond TaAs. Crucially, our results show that Ta$_3$S$_2$ has the largest $k$-space separation between Weyl nodes among known Weyl semimetal candidates, which is about twice larger than TaAs and twenty times larger than WTe$_2$. Moreover, we predict that increasing the lattice by $<4\%$ can annihilate all Weyl nodes, driving the system to a topological insulator state. The robust Weyl semimetal state and the topological metal-to-insulator transition in Ta$_3$S$_2$ are potentially useful in device-applications. Our methodology can be generally applied to search for new robust Weyl semimetals.
\end{abstract}

\maketitle

The rich correspondence between high-energy particle physics and low-energy condensed matter physics has been a constant source of inspiration throughout the history of modern physics \cite{Wilczek}. This has led to important breakthroughs in many aspects of fundamental physics such as the Planck constant and the blackbody radiation, the Pauli exclusion principle and magnetism, as well as the Anderson-Higgs mechanism and superconductivity, which in turn helped us understand materials leading to important practical applications. Recently, there has been significant interest in realizing high energy particles in solid state crystals. The discovery of massless Dirac fermions in graphene and on the surface of topological insulators has taken center stage of the past decade's research in condensed matter and materials science \cite{Graphene, Hasan2010, Qi2011, Hasan2011}. Weyl semimetals \cite{Weyl, Herring, Abrikosov, Ninomiya, Murakami2007, Volovik2003, Wan2011, Ran, Balents_viewpoint, Burkov2011, Hasan_Na3Bi,Huang2015,Weng2015, Hasan_TaAs, MIT_Weyl,TaAs_Ding} are crystals whose quasiparticle excitation is the Weyl fermion \cite{Weyl}, a particle that played a crucial role in the development of quantum field theory and the Standard Model but has not yet been observed as a fundamental particle in Nature. Weyl fermions have a definite left- or right-handed chirality, and can be combined in pairs of opposite chirality to generate a massless Dirac fermion. In a Weyl semimetal, the chirality associated with each Weyl node can be understood as a topologically protected charge, hence broadening the classification of topological phases of matter beyond insulators. The existence of Weyl fermions and the nontrivial topology give rise to a wide range of new fundamental physics, which can be potentially used in future applications. The presence of parallel electric and magnetic fields can break the apparent conservation of the chiral charge, which results in the condensed matter version of the chiral anomaly, making a Weyl semimetal, unlike ordinary nonmagnetic metals, more conductive with an increasing magnetic field \cite{Chiral_anomaly_ChenGF, Chiral_anomaly_Jia}. Remarkably, Weyl nodes are robust against imperfections in the crystal, and are protected by  the crystals inherent translational invariance \cite{Wan2011}, giving rise to an exceptionally high electron mobility \cite{Semimetal_Jia}. These properties suggest that Weyl semimetals may be used to improve electronics by carrying electric currents more efficiently. Because Weyl fermion quasiparticles are naturally spin-momentum locked \cite{Wan2011,Balents_viewpoint, Burkov2011} and superconductivity in these materials may exhibit non-Abelian statistics \cite{Weyl-SC-2,Weyl-SC-5, Weyl-SC-7}, they may also be exploited to realize new applications, such as in spintronics and quantum computers. Furthermore, a monolayer (the 2D limit) of Weyl semimetals can host quantized anomalous Hall (or spin Hall) current without an external magnetic field. In order to make these novel phenomena experimentally accessible, especially at ambient conditions so that they can be utilized in device applications, a robust Weyl semimetal with well-separated Weyl nodes is critically needed.

Recently, the first Weyl semimetal was discovered in the TaAs family \cite{Huang2015,Weng2015, Hasan_TaAs,TaAs_Ding,TaAs_Ding_2,TaAs_Chen,NbAs_Hasan,TaP_Shi,TaP_Hasan,NbP_Hasan,NbP_DLFeng,NbP_Ando, NbP_Chen, NbP_STM}. However, research progress is still significantly held back due to the lack of robust materials candidates. In a Weyl semimetal, Weyl nodes of opposite chirality are separated in momentum space. The degree of separation between Weyl nodes provides a measure of the Weyl phase's ``topological strength'' \cite{MoWTe2} that one has to overcome in order to annihilate the Weyl fermions in pairs. A large $k$ space separation of the Weyl nodes gurantees a robust and stable Weyl semimetal state, which is the prerequisite for observing the many exotic phenomena predicted to be detectable in spectroscopic and transport experiments. Therefore, it is of critical importance to find robust and ideal Weyl semimetals, which have fewer Weyl nodes and more importantly whose Weyl nodes are well separated in momentum space and are located close to the chemical potential in energy.

\begin{figure*}
\includegraphics[width=12cm]{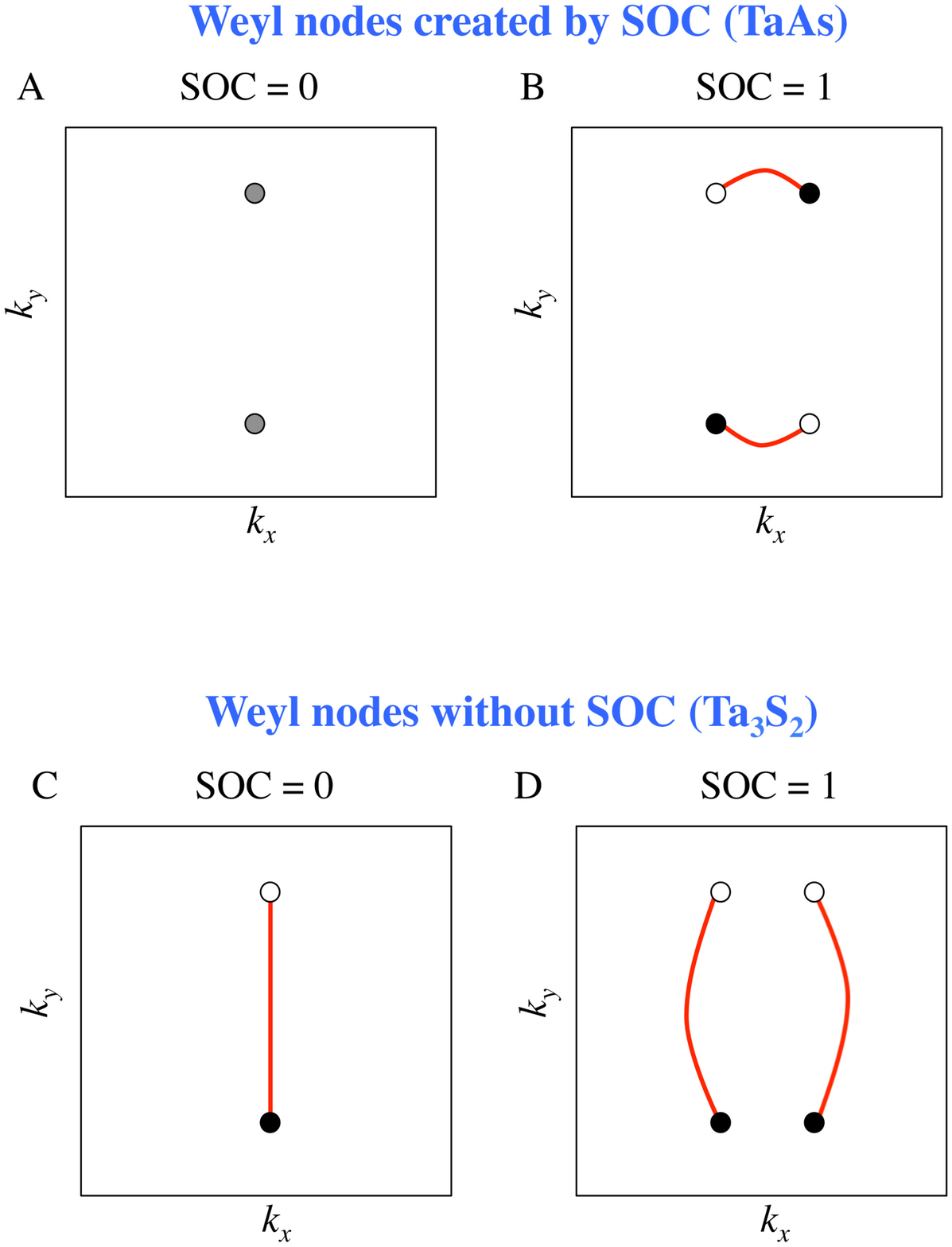}
\caption{
{\bf Methodology to design robust Weyl semimetals with well-separated Weyl nodes.} ({\bf A} and {\bf B}) The previous way of looking for Weyl semimetal states. Without SOC, the conduction and valence bands show some nodal crossings (which are not Weyl nodes). The inclusion of SOC splits each nodal point into a pair of Weyl nodes of opposite chiralities. In this way, the separation between the Weyl nodes is entirely determined by the SOC strength of the compound. For example, the first and the only Weyl semimetal in experiments, TaAs, belongs to this type \cite{Huang2015, Hasan_TaAs}. In the absence of SOC, the conduction and valence bands cross each other to form nodal lines. The inclusion of SOC gaps out the nodal lines and gives rise to 12 pairs of Weyl nodes \cite{Hasan_TaAs}. ({\bf C} and {\bf D}) Our new methodology to design robust Weyl semimetals with well-separated Weyl nodes. We propose to look for compounds that are already a Weyl semimetal without SOC. The inclusion of SOC will split each Weyl node to two nodes of the same chirality. In this way, SOC becomes irrelevant. The separation between the Weyl nodes of opposite chiralities is determined by the magnitude of the band inversion, which is not limited by the SOC strength and can be very large.}
\label{Fig0}
\end{figure*}

We start by describing our methodology to design robust Weyl semimetals with well-separated Weyl nodes. It has been widely accepted that strong spin-orbit coupling (SOC) is a key ingredient for realizing topological states. Our methodology evades this commonly accepted point of view. We show that, to design robust Weyl semimetals with well-separated Weyl nodes, one needs to look for materials (1) that break space inversion symmetry, (2) that have small density of states at the chemical potential, and (3) that are already a Weyl semimetal in the absence of SOC. SOC, on the other hand, does not play a significant role in the whole consideration. We elaborate on our methodology in Fig.~\ref{Fig0}. Figs.~\ref{Fig0}A and B show the previous way of looking for Weyl semimetals. Specifically, without SOC, the conduction and valence bands show some nodal crossings (which are not Weyl nodes). The inclusion of SOC splits each nodal point into a pair of Weyl nodes of opposite chiralities. In this way, the separation between the Weyl nodes is entirely determined by the SOC strength of the compound. For example, the first and the only Weyl semimetal in experiments, TaAs, belongs to this type \cite{Huang2015, Hasan_TaAs}. It is worth noting that TaAs has almost the strongest SOC that one could achieve in real materials. Even then, the separation was only barely resolved in experiments \cite{Hasan_TaAs}. Finding Weyl semimetals with larger separation than TaAs using this previous method is not possible. By contrast, in Figs.~\ref{Fig0}C and D, we present a new methodology. We propose to look for compounds that are already a Weyl semimetal without SOC (Figs.~\ref{Fig0}C). The inclusion of SOC will split each Weyl node to two nodes of the same chirality. In this way, SOC becomes irrelevant. The separation between the Weyl nodes of opposite chiralities is determined by the magnitude of the band inversion, which is not limited by the SOC strength and can be very large. Therefore, our new methodology can give rise to robust Weyl semimetals with well-separated Weyl nodes.

\begin{figure*}
\includegraphics[width=17cm]{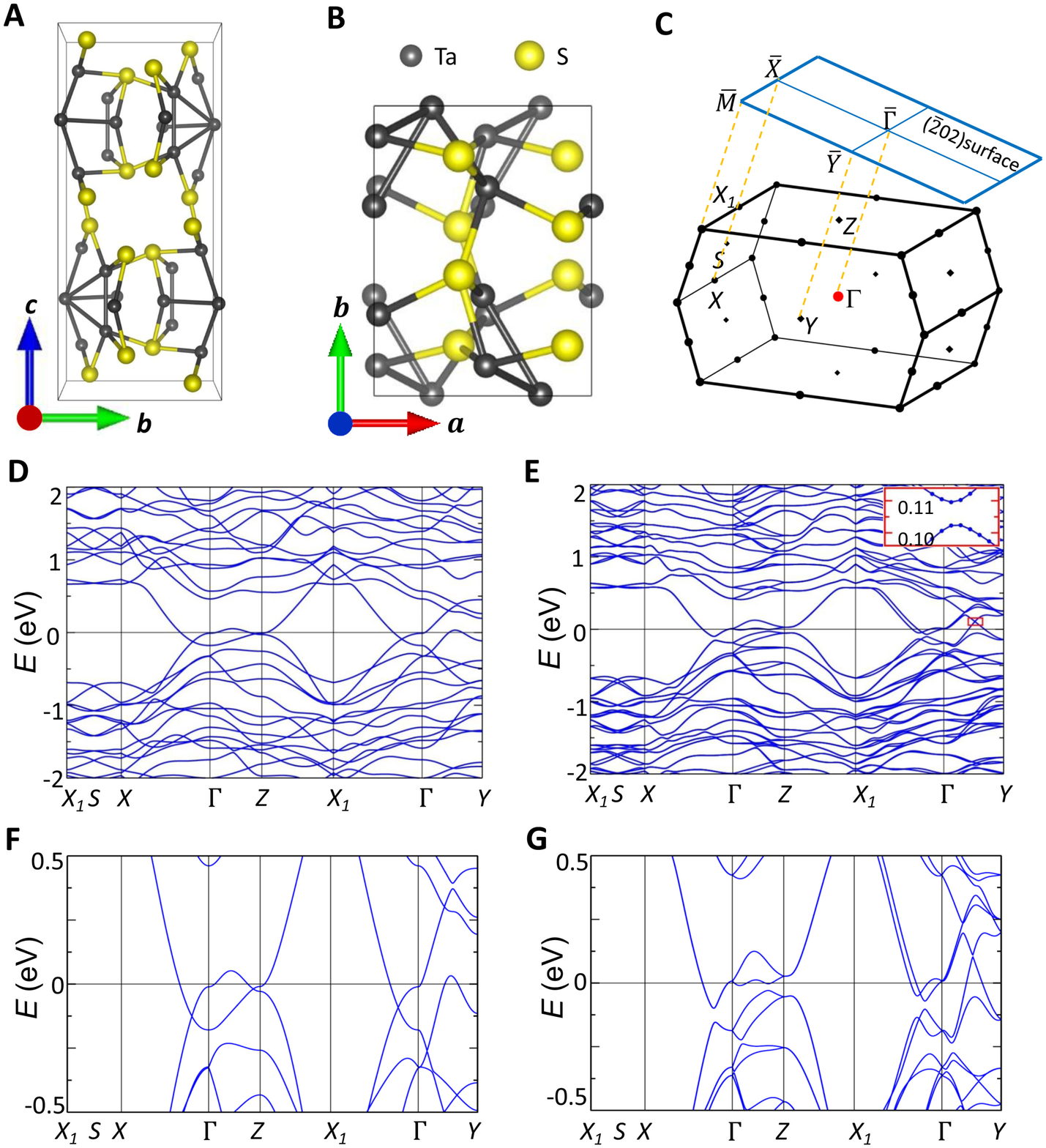}
\caption{
{\bf Crystal and band structure of Ta$_3$S$_2$.} {\bf A,} Side view of the crystal structure for one unit cell of Ta$_3$S$_2$. The dark-grey and yellow balls represent the Ta and S atoms, respectively. {\bf B,} Same as (A) but for the top view. Ta$_3$S$_2$ crystallizes in a base-centered orthorhombic structure with space group $Abm2$ ($\#39$) with lattice constants $a=5.6051 \textrm{\AA}$, $b=7.4783 \textrm{\AA}$ and $c=17.222 \textrm{\AA}$. As can be seen from the crystal lattice structure, Ta$_3$S$_2$ lacks inversion symmetry.}
\label{Fig1}
\end{figure*}

Using this methodology as a guideline, we have searched among the material parameter space and identified by far the most robust and ideal Weyl semimetal candidate in the inversion breaking, single crystalline compound tantalum sulfide Ta$_3$S$_2$ with new and novel properties beyond TaAs. Ta$_3$S$_2$ crystalizes in a base-centered orthorhombic structure \cite{Ta3S2_Crystal_1, Ta3S2_Crystal_2}. Single crystals of this compound has been grown \cite{Ta3S2_Crystal_1, Ta3S2_Crystal_2}, and transport experiments indeed reported a semimetallic behavior \cite{Ta3S2_Crystal_2}. The lattice constants are $a=5.6051 \textrm{\AA}$, $b=7.4783 \textrm{\AA}$ and $c=17.222 \textrm{\AA}$, and the space group $Abm2$ ($\# 39$). There are 24 Ta atoms and 16 S atoms in a conventional unit cell (Figs.~\ref{Fig1}A,B). It can be seen that the lattice lacks space-inversion symmetry, which is key to realizing the Weyl semimetal state in this time-reversal invariant system. Moreover, the system has two glide mirror symmetries associated with the $y$ and $z$ directions, i.e., $\mathcal{G}_y: (x,y,z) \Rightarrow (x,-y,z+0.5)$ and $\mathcal{G}_z: (x,y,z) \Rightarrow (x,y,-z+0.5)$, but it does not have any mirror symmetry along the $x$ direction. The symmetry condition determines the number, energy and momentum space configuration of the Weyl nodes in Ta$_3$S$_2$, which will be discussed below.

Figs.\ref{Fig1}D,F shows the first-principles calculated band structure in the absence of spin-orbit coupling (SOC), from which it can be seen that the conduction and valence bands dip into each other, suggesting a semimetallic ground state. Particularly, we find that the conduction and valence bands cross each other without opening up a gap along the $\Gamma-X-Z-X_1$ direction. Upon the inclusion of SOC (Figs.\ref{Fig1}E,G), the band structure is found to be fully gapped along all high symmetry directions.

In order to search for the Weyl nodes in Ta$_3$S$_2$, we calculated the band structure throughout its Brillouin zone (BZ). In the absence of SOC (Fig.\ref{Fig2}A), we found a line node on the $k_y=0$ plane, which is the band crossings along $\Gamma-X-Z-X_1$ direction as seen in Fig.\ref{Fig1}D. We note that this line node is on the $k_y=0$ plane and is,therefore, protected by the mirror symmetry $\mathcal{G}_y$. In addition, we also found two pairs of Weyl nodes located on the $k_x=0$ plane (Fig.\ref{Fig2}A). We determine the chirality of the Weyl node by computing the Berry curvature through a closed 2D manifold enclosing the node. Considering the available symmetries as discussed above, Ta$_3$S$_2$ has only one irreducible pair of Weyl nodes. The second pair is obtained by applying the Mirror operation $\mathcal{G}_z$. Hence the separation between the pair of Weyl nodes of opposite chiralities in Ta$_3$S$_2$ is irrelevant to SOC. It is rather determined by the magnitude of the inversion between the lowest conduction and valence bands. Upon the inclusion of SOC, each Weyl node without SOC splits into two spinful Weyl nodes of the same chirality. This is quite intuitive because each state without SOC should be considered as two states of opposite spins. For this reason there are four pairs of Weyl nodes in the presence of SOC. Also because of the mirror symmetries, all the Weyl nodes have the same energy in Ta$_3$S$_2$. We show the dispersion away from a Weyl node along all three momentum space directions in Figs.\ref{Fig2}C,D. It can be seen that Ta$_3$S$_2$'s Weyl nodes are of the so-called type II \cite{WTe2} because the two bands that cross to form the Weyl nodes have the same sign of velocity along one momentum direction (in this case, $k_y$). In the presence of SOC (Fig.\ref{Fig2}D), the Weyl nodes are approximately $10$ meV below the Fermi level, which is in contrast to the case in the Mo$_x$W$_{1-x}$Te$_2$ systems \cite{WTe2, MoWTe2}. This makes Ta$_3$S$_2$ more hopeful than Mo$_x$W$_{1-x}$Te$_2$ \cite{WTe2, MoWTe2} for observing the type II Weyl nodes by photoemission experiments. More importantly, the $k$ separation of the Weyl nodes (Fig.\ref{Fig2}D) in Ta$_3$S$_2$ is as large as $\sim0.15$ $\textrm{\AA}^{-1}$. This is by far the largest in known Weyl semimetal candidates, and, in fact, twice larger than the measured value in TaAs (about $0.07$ to $0.08$ $\textrm{\AA}^{-1}$) \cite{Hasan_TaAs} and twenty times larger than the predicted value in WTe$_2$ ($\sim0.007$ $\textrm{\AA}^{-1}$) \cite{WTe2}. The facts that the Weyl nodes are well-separated in momentum space and that they are located close to the chemical potential make Ta$_3$S$_2$ the by far most robust and ideal Weyl semimetal candidate for observing and realizing the novel Weyl physics in both spectroscopic and transport experiments.

Another signature of the Weyl semimetal state is the Fermi arc electron states on the surface of the crystal. Figure~\ref{Fig3} shows the calculated surface state electronic structure on Ta$_3$S$_2$'s $(\bar202)$ surface. The calculated surface Fermi surface (Figs.~\ref{Fig3}A,D) show a rich structure, including both topological Fermi arcs and topologically-trivial surface states. There are finite projected bulk Fermi surfaces as shown by the shaded areas in Fig.~\ref{Fig3}D, because the all Weyl nodes are type II. It is known that, at the energy of a type II Weyl node, the bulk Fermi surface is not an isolated point but a touching point between an electron and a hole pockets \cite{WTe2}. It is interesting to note that the surface Fermi surface does not respect mirror symmetry along the $\bar{\Gamma}-\bar{X}$ or $\bar{\Gamma}-\bar{Y}$ axis. As seen from Fig.~\ref{Fig1}B, the $\bar{\Gamma}-\bar{Y}$ is the projection of the $k_x=0$ plane, which is not a mirror plane. The $\bar{\Gamma}-\bar{X}$ is the projection of the $k_y=0$ plane, which is indeed a mirror plane. However, it corresponds to a glide mirror operation, $\mathcal{G}_y: (x,y,z) \Rightarrow (x,-y,z+0.5)$. Therefore, the $(\bar202)$ surface breaks the glide mirror symmetry. To visualize the Fermi arc surface states, Fig.~\ref{Fig3}B shows the energy dispersion cut along the $Cut 1$ (defined by the red dotted line in Fig.~\ref{Fig3}A). In both box 1 and box 2, we clearly see that a surface state is terminated directly onto a Weyl node, which is the touching point between the shaded areas. This calculation demonstrates the existence of the Fermi arc surface states. Specifically, we label the two surface states in the box 1 (upper right panel in Fig.~\ref{Fig3}B) as $\beta$ and $\alpha$ from left to right. We see that $\beta$ is the Fermi arc. In Fig.~\ref{Fig3}D, we show the high-resolution Fermi surface zoomed in near the $\bar{X}$ point. The Weyl node that corresponds to the box 1 is the black dot that is directly above the the bottom-left most black dot in Fig.~\ref{Fig3}D. The two surface states, $\alpha$ and $\beta$ are be identified, and, indeed, the left surface state, $\beta$, is the Fermi arc terminating onto this Weyl node. Through similar analyses, we can determine that $\alpha$ is the Fermi arc corresponding to the bottom-left most Weyl node. Based on the above analyses, we show the determined Fermi arc connection in Fig.~\ref{Fig3}E. We also emphasize that the topological band theory of the Weyl semimetal phase only requires the number of Fermi arcs that are terminated on a given projected Weyl node must equal the absolute value of its chiral charge. The detailed connectivity pattern can vary based on the surface conditions such as the surface potential, surface relaxation, surface density, etc. Hence the purpose of Fig.~\ref{Fig3} is to show the existence of the Fermi arcs, which is a key signature of the Weyl semimetal state in Ta$_3$S$_2$. The details of the surface electronic structure, including the connectivity pattern, will depend on the surface conditions, which have to be determined by experiments.

Now we show the topological metal to insulator transition in Ta$_3$S$_2$ in Fig.~\ref{Fig4}. In order to best visualize the transition, we show the band dispersion along a cut that goes through an irreducible pair of Weyl node (as defined by the red dotted line in Fig.~\ref{Fig5}D). As shown in Fig.~\ref{Fig4}A, we show this cut as a function of different values of the lattice constant $b$. At the original lattice constant, $b'=b$, we indeed see a pair of type II Weyl nodes, as expected. As we increase the lattice constant by $3.0\%$, $b'=1.030b$, we see that Weyl nodes approach each other and their separation decreases by half. As we further increase the lattice constant to $b'=1.040b$, the two Weyl nodes annihilate with each other and the band structure become fully gapped. By a careful calculation, we determined that the lattice constant corresponding to the critical point is $b'=1.037b$. The resulting fully gapped state for $b'>1.037b$ has two possible fates, i.e., either being a trivial insulator or being a topological insulator. We have calculated the the Wilson loop of the Wannier function centered on the $k_z= 0$ plane and on the $k_z = \pi$ plane (Fig.~\ref{Fig4}B), from which we determined that the gapped state for $b'>1.037b$ is a topological insulator and its $\mathcal{Z}_2$ indices are ($1;000$). Therefore, by increasing the lattice constant $b$, one can realize a topological phase transition from a Weyl semimetal state to a topological insulator state in Ta$_3$S$_2$. The corresponding evolution of the surface electronic structure is shown in Fig.~\ref{Fig4}C for the case of the $(\bar202)$ surface. Interestingly, the projected Weyl nodes of opposite chirality, which are connected by the Fermi arcs, approach each other and eventually meet on the $\bar{X}-\bar{\Gamma}$ axis where they annihilate. The resulting surface has a single surface state whose Fermi surface encloses the Kramers' point $\bar{X}$, which also demonstrates the topological insulator state.

Besides the topological phase transition, we found that the system exhibits other important tunabilities. Specifically, it can be seen that while the Weyl nodes at the original lattice constant are type II, they become type I at $b'=1.030b$ because the Weyl node is now formed by two bands with the opposite sign of velocity (Fig.~\ref{Fig4}A). Therefore, interestingly, there is a transition from type II Weyl fermions to type I Weyl femions as one increases the lattice constant $b$. Moreover, we note that the energy of the bands in Fig.~\ref{Fig4}A shifts across the chemical potential as one increases $b$. Therefore, there also exist specific values of the lattice constant $b$, at which important features will be moved exactly onto the chemical potential. We focuses on two important features, i.e., the Weyl nodes and the van Hove singularities (VHS) that arise from the Weyl cones (see middle panel of Fig.~\ref{Fig4}A). Placing the Weyl nodes at the Fermi level is very meaningful because the Weyl nodes are monopoles of Berry curvature. Thus any novel phenomenon that arises from the chirality of the Weyl fermions such as the chiral anomaly will become the most significant when the Weyl nodes are at the Fermi level. Putting the VHS at the Fermi level can also interesting because the VHS is due to a saddle point in the band structure, which means that the density of states will show a maxima at the energy of the VHS. An enhanced density of states is favorable for inducing correlated physics such as superconductivity or magnetism. A detailed phase diagram is shown in Fig.~\ref{Fig4}C.

We show that band structure of Ta$_3$S$_2$ exhibits a new type of critical point as one decreases the lattice constant $c$ or increases the spin-orbit coupling $\lambda$. As discussed above, at ambient conditions, the conduction and valence bands only touch each other at 8 discrete points in the BZ, which are the 8 Weyl nodes. Here, we show that decreasing the lattice constant $c$ or increasing the spin-orbit coupling $\lambda$ leads to the generation of new Weyl nodes. Representing the critical point for this process is the critical value for the lattice constant $c$ or spin-orbit coupling $\lambda$ corresponding to the point where the conduction and valence bands just touch, $c_{\textrm{critical}}=0.98c$ or $\lambda_{\textrm{critical}}=1.027\lambda$. Taking the critical point of $\lambda_{\textrm{critical}}=1.027\lambda$ as the example, we show the $k$ space locations of these newly emerged band touchings by the green dots in Fig.~\ref{Fig5}D. Interestingly, we find that the critical point band structure is novel. Specifically, the dispersion along $k_z$ near the band touching behaves like two downward facing parabolas. These two parabolas touch at their vertex, which forms the band touching point. This is distinct from the critical points associated with any previously known Weyl semimetal candidates. For example, the critical point band structure of TaAs can be thought of as two parabolas of opposite directions, one facing up and the other facing down (Fig.~\ref{Fig5}A). Then entering the Weyl phase from the critical point essentially means ``pushing'' the two parabolas ``into'' each other so that they cross to form the two Weyl nodes. The situation in the Mo$_x$W$_{1-x}$Te$_2$ system is very similar, but the only difference is that the direction of the parabolas is titled away from being vertical (Fig.~\ref{Fig5}B). By contrast, in Ta$_3$S$_2$ we have two parabolas that face the same direction (Fig.~\ref{Fig5}C). A distinct and unique property of the new critical point is that it leads to a saddle point in the band structure, giving rise to a van Hove singularity. The saddle point behavior can be seen from the band dispersions shown in Fig.~\ref{Fig5}E. If one focuses on the conduction band in Fig.~\ref{Fig5}E, the touching point is the energy minima for the dispersions along the $k_x, k_y$ directions but it is the energy maxima along the $k_z$ direction. The saddle point band structure brings about a van Hove singularity, which generates a maxima in the density of states (DOS) and a divergence in the first derivative of the DOS at the energy of the van Hove singularity, as seen in Fig.~\ref{Fig5}G.

We elaborate on the meaning of the \textbf{robust} Weyl semimetal candidate as emphasized in our paper. Firstly, we mean that the realization of the candidate is likely to be experimentally feasible. This involves the following critical conditions: (1) The prediction is based on the realistic crystal structure, meaning that the compound does crystalize in the proposed crystal structure at ambient conditions. (2) The prediction does not require fine tuning of the chemical composition or the magnetic domains. (3) The Weyl nodes are not located at energies far above the chemical potential so that they can be observed by photoemission. This was the case for our prediction of TaAs \cite{Huang2015}, which has now been realized \cite{Hasan_TaAs}. This is also the case here for Ta$_3$S$_2$, which demonstrates that the experimental feasibility of our proposal on Ta$_3$S$_2$. Secondly, and more importantly, the term ``robust'' refers to a large separation of the Weyl nodes in momentum space because, as discussed above, the separation of the Weyl nodes provides a measure of a Weyl semimetal's topological strength. We highlight again that Ta$_3$S$_2$ has the largest $k$-space separation between Weyl nodes among known Weyl semimetal candidates, which is about twice larger than TaAs. This will greatly help resolve the Weyl nodes in various spectroscopic measurements such as photoemission and scanning tunneling spectroscopies. This will also make it easier to probe the chiral anomaly and other monopole physics in electrical and optical transport experiments.

Finally, we compare the topological metal-to-insulator transition in Ta$_3$S$_2$ with transitions predicted in other Weyl candidates \cite{Vanderbilt, BiX}. A theoretical work \cite{Vanderbilt} predicted the topological phase transitions from a trivial band insulator to a Weyl semimetal and then to a topological insulator by varying the chemical composition $x$ in LaBi$_{1-x}$Sb$_x$Te$_3$ or applying external pressure to BiTeI. However, the composition or pressure range that corresponds to Weyl semimetal phase is predicted to be extremely narrow \cite{Vanderbilt}. Hence it requires ultra-fine tuning, which is very difficult in experiments. Also, LaBi$_{1-x}$Sb$_x$Te$_3$ has never been grown in the crystal structure required by the proposal \cite{Vanderbilt} at least in the single crystal form. Another work \cite{BiX} proposed similar transitions in the $\beta$-Bi$_4$Br$_4$ under external pressure. In order to induce a Weyl semimetal phase in the $\beta$-Bi$_4$Br$_4$ crystal structure that does have inversion symmetry, a hypothetical inversion breaking term was assumed in the calculation. By contrast, Ta$_3$S$_2$ is an inversion breaking, single crystalline compound. Single crystalline Ta$_3$S$_2$ samples have been grown \cite{Ta3S2_Crystal_1, Ta3S2_Crystal_2}. The Weyl semimetal state is stable and does not require fine tuning. We propose the following three methods for increasing the $b$ lattice constant: (1) It can by done by applying external force. The method has been demonstrated in Ref. \cite{Zheng}. As an order-of-magnitude estimate, we calculated the required force by first-principles calculations and obtained a $\sim6$ GPa force for a $\sim4\%$ increase of the $b$ lattice constant. Force in this range is experimentally feasible and the change of the lattice constant can be monitored by a transmission electron microscopy \cite{Zheng}. (2) It may also be achieved by growing Ta$_3$S$_2$ film onto a substrate with lattice mismatch. (3) It may be achieved by growing samples with isoelectronic chemical substitution such as Ta$_3$(S$_{1-x}$Se$_x$)$_2$. These facts highlight that Ta$_3$S$_2$ is, to date, not only the most ideal platform for advancing our understanding of Weyl semimetals and Weyl physics, but to also facilitate the exploitation of the exotic and novel properties in future device applications.

\clearpage

\clearpage
\addtocounter{figure}{-1}
\begin{figure*}[t!]
\caption{{\bf C,} The primitive first Brillouin zone of Ta$_3$S$_2$ showing a base-centered orthorhombic structure and the location of relevant high-symmetry points. Representing the ($\bar{2}02$) surface is a blue rectangular plane. The yellow dashed lines represent the projection of high-symmetry points on the ($\bar{2}02$) surface. {\bf D,} First-principles electronic band structure calculation for Ta$_3$S$_2$ with the absence of spin-orbit coupling. Bulk band crossings observed along the $\Gamma$-X-Z-X$_1$ direction line near the Fermi level are a part of a line node on the $k_y=0$ plane. {\bf E,} Same as (D) but with the addition of spin-orbit coupling. After the incorporation of spin-orbit coupling, the band structure is fully gapped along all high symmetry directions. Zooming into the red box along the $\Gamma$-Y direction shows a minimum gap size between the conduction band minimum and valence band maximum of $\approx{7}$ meV, which is 96.5\% smaller than the 200 meV gap observed along the same direction in (D). ({\bf F}-{\bf G,}) Zooming in close to the Fermi level for (D)-(E), respectively. Upon the inclusion of SOC, we clearly observe that band structure becomes fully gapped along along all high symmetry directions}
\end{figure*}

\begin{figure*}[t]
\includegraphics[width=16cm]{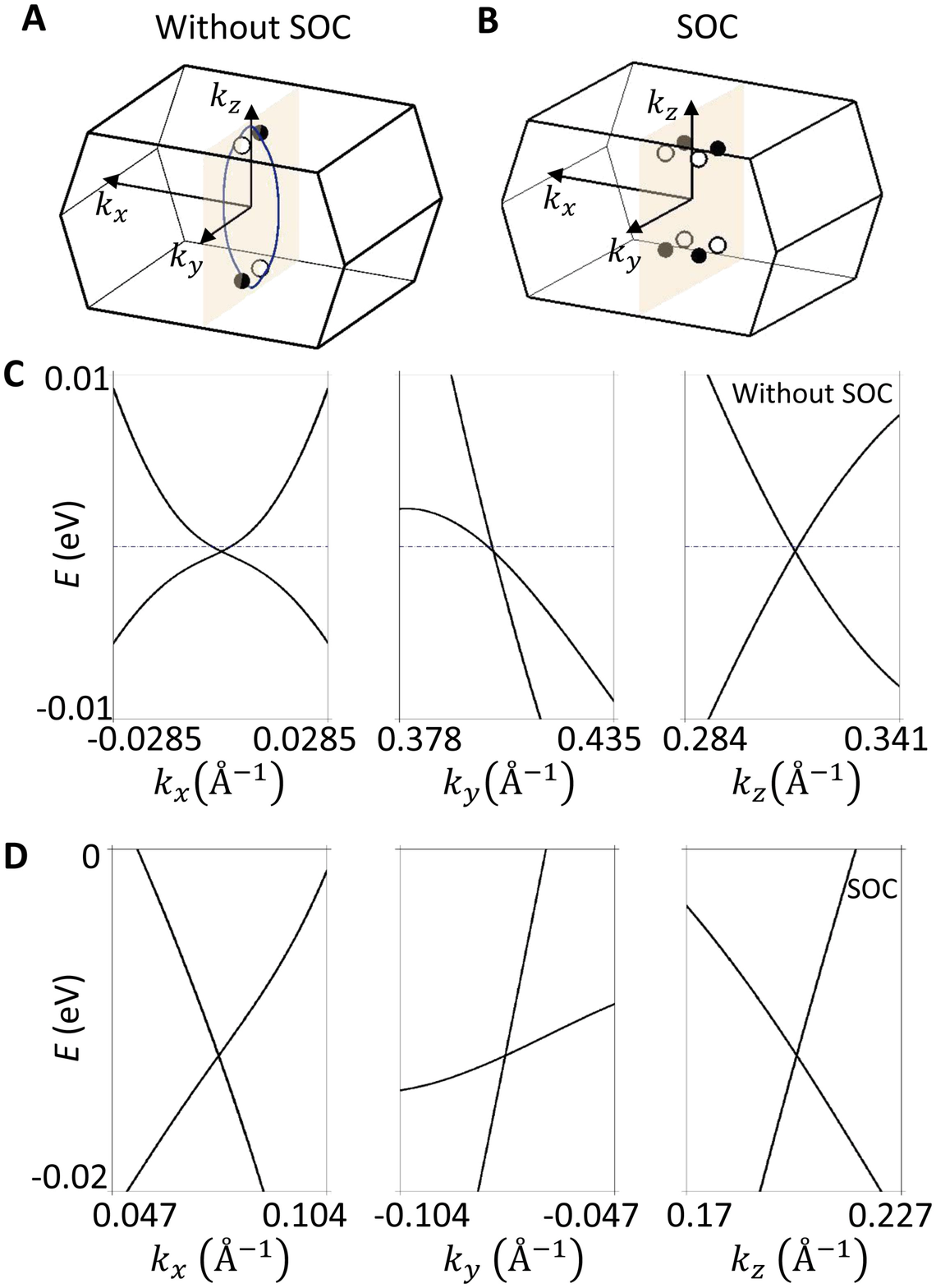}
\caption{
{\bf Weyl nodes in Ta$_3$S$_2$.} ({\bf A}) The primitive first Brillouin zone (BZ) of Ta$_3$S$_2$ showing the $k$-space location of the Weyl points in the absence of spin-orbit coupling.}
\label{Fig2}
\end{figure*}
\addtocounter{figure}{-1}
\begin{figure*}[t!]
\caption{The black and white circles represent Weyl points with opposite chirality. On the $k_x=0$ plane, represented in a tan color, we observe two pairs Weyl nodes with chiral charge $\pm{1}$. Only one pair of Weyl nodes is an irreducible representation. Shown in blue is the line node on the $k_y=0$ plane that corresponds to the crossing along the $\Gamma$-X-Z-X$_1$ direction. ({\bf B}) Same as (A) but after the incorporation of spin-orbit coupling. Each Weyl node splits into two spinful Weyl nodes of the same chirality and moves into the bulk. Therefore, we observe 8 Weyl nodes in a single BZ. ({\bf C}) The energy dispersion along the three momentum space directions of the Weyl node in the absence of spin-orbit coupling. ({\bf D}) Same as panel C but with the addition of spin-orbit coupling. From this series of energy dispersion cuts along the three momentum space directions around the Weyl point reveals its type II character (i.e. observe in along $k_y$ that both crossing bands have the same sign of velocity along one momentum direction. Furthermore, observe that the Weyl nodes also move deeper in binding energy, $\approx{-10}$ meV.}
\end{figure*}

\begin{figure*}[t]
\includegraphics[width=17cm]{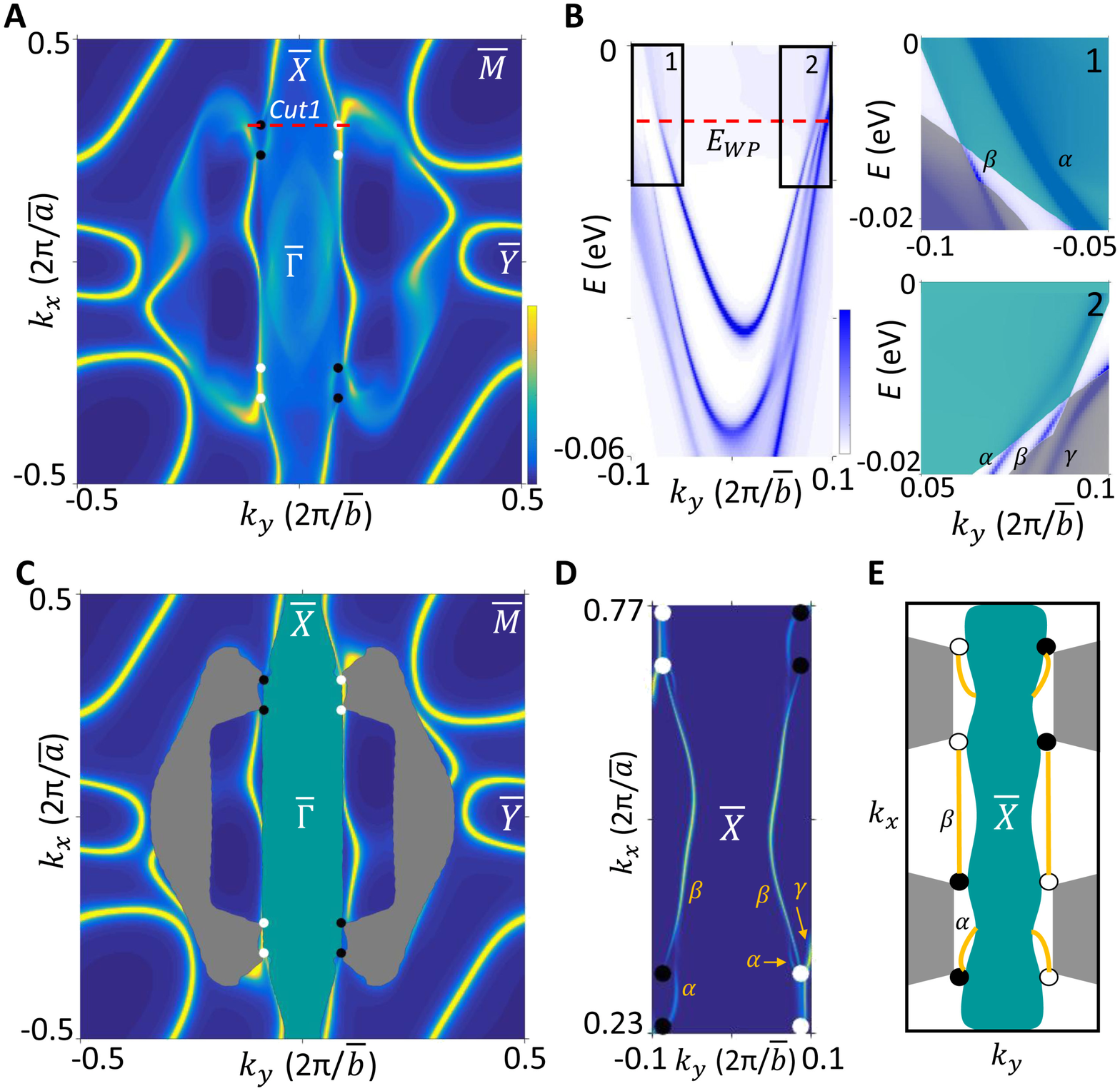}
\caption{ {\bf Fermi arcs in Ta$_3$S$_2$.} ({\bf A}) Surface state spectrum for the ($\bar{2}02$) surface of Ta$_3$S$_2$ at constant energy. Shown in yellow are the surface states. A finite amount of projected bulk Fermi surfaces is observed due to the type II character of the Weyl nodes. Denoted by black and white circles are the projected Weyl nodes. ({\bf B}) Energy dispersion cut along Cut1 (red dashed line) for a pair of projected Weyl nodes around $\bar{X}$ in (A). As a guide, the red dashed line represents the energy level of the Weyl node (E$_{W1}\approx{-10}$ meV). Left panel shows two surface states (blue curves) emerging from box1 and three entering box2. Top-rightmost panel is zoomed into box1 and shows a touching point, which defines the Weyl node as type II, and two Fermi arc surface states that are labeled $\alpha$ and $\beta$. The $\beta$ Fermi arc surface state terminates onto the Weyl node.}
\label{Fig3}
\end{figure*}
\addtocounter{figure}{-1}
\begin{figure*}[t!]
\caption{({\bf B}) Bottom-rightmost panel is similar to the top-right most panel. Here we observe three Fermi arc surface states, $\alpha$, $\beta$, and $\gamma$. Again, the $\beta$ Fermi arc surface state is observed to terminate onto the Weyl node. ({\bf C}) Same as (a) but with the projected bulk bands represented by shaded areas. The Fermi arc surface states connecting pairs of projected Weyl nodes are distinguishable from the remaining trivial surface states. ({\bf D}) Zoom-in of the surface states around the projected Weyl nodes near $\bar{X}$. Shown, among other things, are 8 projected Weyl nodes. Single Fermi arc surface states, $\beta$, are shown to connect two individual pairs of Weyl nodes, which is consistent with the predicted $\pm{1}$ chiral charge for each Weyl node. The $\alpha$ and $\gamma$ Fermi arc surface states described in (B) are also labeled. ({\bf E}) Cartoon illustration of (D) showing a possible Fermi arc linking scheme. The exact connectivity pattern of Fermi arcs with Weyl nodes depends on the surface conditions.}
\end{figure*}

\begin{figure*}[t]
\includegraphics[width=17cm]{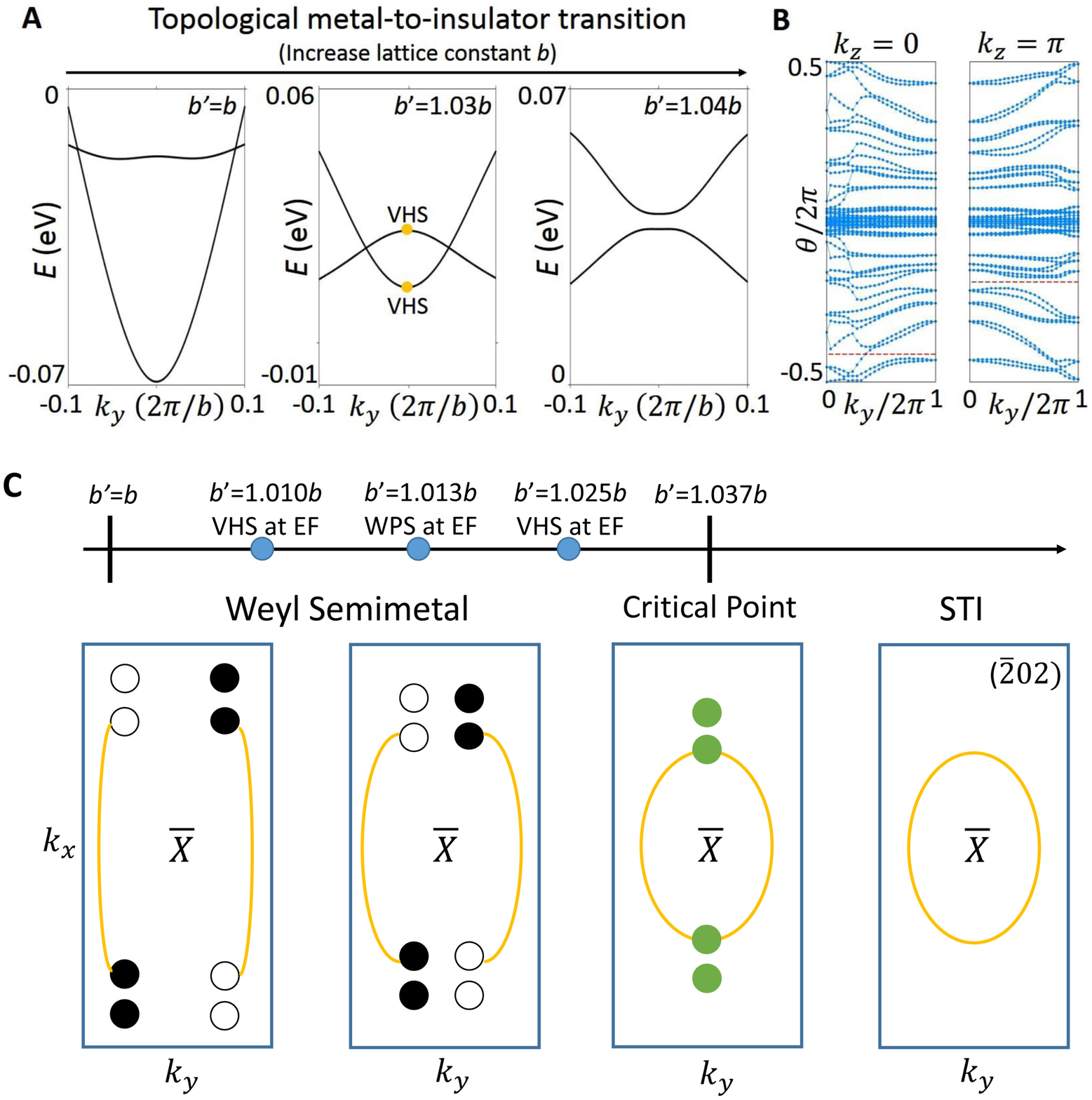}
\caption{ {\bf Topological metal-to-insulator transition and tunabilities.} ({\bf A}) Energy dispersion cut along $k_y$ for an irreducible pair of Weyl nodes as the $b$ lattice constant is tuned to various values. At the $b'=b$ (left panel) we observe a pair of type II Weyl nodes. At $b'=1.03b$ (middle panel) the two Weyl nodes approach each other and their separation decreases by about half. Another important feature shown here is the van Hove singularities (VHS), which are labeled with yellow circles. At $b'\geq{1.04b}$ (right panel) the two Weyl nodes annihilate each other and the band structure becomes fully gapped. ({\bf B}) Calculation of the Wilson loop of the Wannier function centered in the $k_z$= 0 plane and in the $k_z$ = $\pi$ plane is performed to confirm the strong topological insulator (STI) state. Red-dashed lines are arbitrary reference positions.}
\label{Fig4}
\end{figure*}
\addtocounter{figure}{-1}
\begin{figure*}[t!]
\caption{The left panel shows one band crossing the red-dashed line, which corresponds to $Z_2$ invariant of 1 for $k_z$=0 plane. The right panel shows no band crossing the red dashed line, which corresponds to $Z_2$ invariant of 0 for $k_z$=$\pi$ plane. $Z_2$=1 confirms the STI state in Ta$_3$S$_2$. ({\bf C}) An illustration of the topological metal to insulator transition for the Weyl nodes around the $\bar{X}$ pocket on the ($\bar{2}02$) surface. Shown in yellow are Fermi arc surface states. As $b'$ is tuned from, the Weyl nodes reach the critical point at $b'=1.037$ and system becomes fully gapped for $b'>1.037$ with a single topological surface state with Fermi surface enclosing the Kramers' point $\bar{X}$, resulting a topological insulator state. Furthermore, as $b'$ is tuned from its original value, the van Hove singularities (VHS) arising from the Weyl cone reach the Fermi level when $b'=1.01b$ and $b'=1.025b$. The Weyl nodes reach the Fermi level when $b'=1.013b$.}
\end{figure*}

\clearpage
\begin{figure*}[t]
\includegraphics[width=17cm]{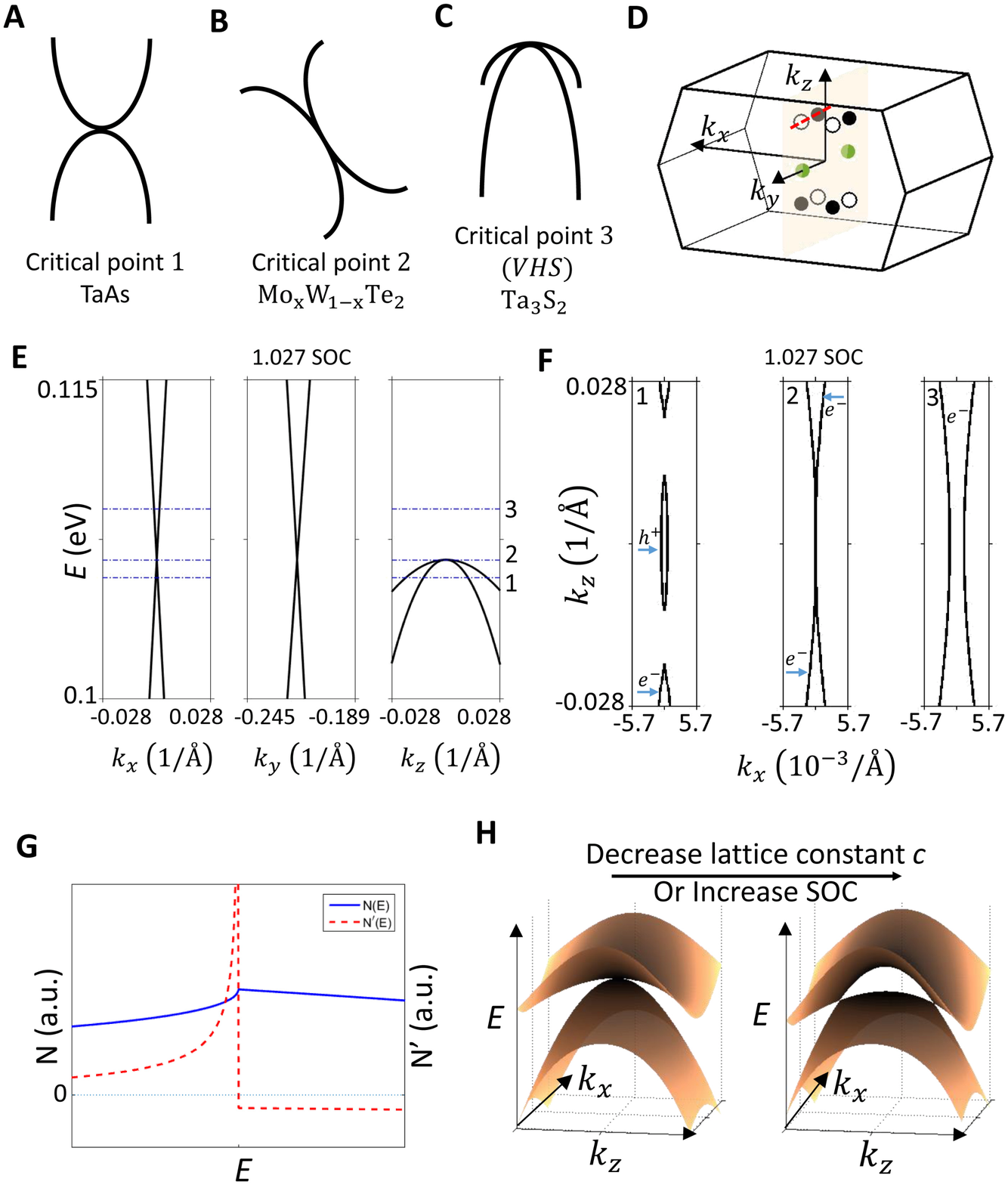}
\caption{ {\bf A new type of critical point.} ({\bf A}) Illustration of the critical point for TaAs as two opposite facing parabolas are driven into each other. ({\bf B}) Similar to (a) but for Mo$_x$W$_{1-x}$Te$_2$, which shows two oppositely facing parabolas that are tilted away from being vertical. ({\bf C}) Ta$_3$S$_2$ shows two parabolas that are facing in the same direction. This new type of critical point leads to a saddle point in the band structure and gives rise to a VHS. ({\bf D}) The momentum space locations of the newly emerged band touchings resulting from driving the system's}

\label{Fig5}
\end{figure*}
\addtocounter{figure}{-1}
\begin{figure*}[t!]
\caption{conduction and valence bands towards each other by changing spin-orbit coupling $\lambda$ to the critical point $\lambda_{critical}= 1.027\lambda$, which are represented by the green dots. ({\bf E}) The saddle point behavior is clearly demonstrated from the band dispersions for $\lambda_{critical}= 1.027\lambda$ by noting that touching point for the conduction band along the $k_x$ and $k_y$ directions is at the energy minimum while being at the energy maximum along the $k_z$ direction. ({\bf F}) The Fermi surfaces corresponding to energy levels 1,2, and 3 along the $k_x$ direction in the far-left panel in (E). ({\bf G}) The saddle point band structure brings about the VHS, which generates a maxima in the density of states (DOS) ($N$, blue solid line) and a divergence in the first derivative of the DOS ($N'$, red dashed line) at the energy of the VHS. ({\bf H}) Illustrations of the conduction and valence bands along the $k_z$ direction as the system is driven to it's critical points by either decreasing the $c$ to $c_{critical}$ lattice constant or increasing the spin-orbit coupling strength $\lambda$ to $\lambda_{critical}$.}
\end{figure*}

\begin{thebibliography}{99}

\bibitem{Wilczek}F. Wilczek,  Why are there Analogies between Condensed Matter and Particle Theory? \textit{Phys. Today} $\mathbf{51}$, 11-13 (1998).
\bibitem{Graphene}A. K. Geim   and  K. S. Novoselov, The rise of graphene. \textit{Nat. Mater.} $\mathbf{6}$, 183-191 (2007).

\bibitem{Hasan2010} M. Z. Hasan and C. L. Kane, Colloquium: Topological insulators. \textit{Rev. Mod. Phys.} \textbf{82}, 3045-3067 (2010).

\bibitem{Qi2011} X.-L. Qi  and  S.-C. Zhang, Topological insulators and superconductors. \textit{Rev. Mod. Phys.} \textbf{83}, 1057-1110 (2011).
\bibitem{Hasan2011} M. Z. Hasan,  and J. E. Moore, Three-Dimensional Topological Insulators. \textit{Ann. Rev. Cond. Mat. Phys.} $\mathbf{2}$, 55-78 (2011).
\bibitem{Weyl} H. Weyl, Elektron und gravitation.  \textit{I. Z. Phys.} $\mathbf{56}$, 330-352 (1929).

\bibitem{Herring} C. Herring, Accidental Degeneracy in the Energy Bands of Crystals. \textit{Phys. Rev.} $\mathbf{52}$, 365-373 (1937).

\bibitem{Abrikosov} A. A. Abrikosov  and S. D. Beneslavskii, Some Properties of Gapless Semiconductors of the Second Kind. \textit{ J. Low Temp. Phys.} $\mathbf{5}$, 141-154 (1972).

\bibitem{Ninomiya} H. B. Nielsen  and  M. Ninomiya, The Adler-Bell-Jackiw anomaly and Weyl fermions in a crystal \textit{Phys. Lett. B} $\mathbf{130}$, 389-396 (1983).
\bibitem{Volovik2003} G. E. Volovik, \textit{The Universe in a Helium Droplet} (Oxford University Press, 2003).

\bibitem{Murakami2007} S.  Murakami, Phase transition between the quantum spin Hall and insulator phases in 3D: emergence of a topological gapless phase. \textit{New J. Phys.} \textbf{9}, 356 (2007).

\bibitem{Wan2011} X. Wan, A. M. Turner, A.  Vishwanath and S. Y. Savrasov, Topological semimetal and Fermi-arc surface states in the electronic structure of pyrochlore iridates. \textit{Phys. Rev. B} \textbf{83}, 205101 (2011).

\bibitem{Ran} K.-Y. Yang, Y.-M. Lu and Y. Ran, Quantum Hall effects in a Weyl semimetal: Possible application in pyrochlore iridates. \textit{Phys. Rev. B} $\mathbf{84}$, 075129 (2011).

\bibitem{Balents_viewpoint} L.  Balents, Weyl electrons kiss. \textit{Physics} \textbf{4}, 36 (2011).

\bibitem{Burkov2011} A. A. Burkov  and L. Balents,  Weyl Semimetal in a Topological Insulator Multilayer. \textit{Phys. Rev. Lett.} \textbf{107}, 127205 (2011).
\bibitem{Hasan_Na3Bi} S.-Y. Xu \textit{et al.}, Observation of Fermi arc surface states in a topological metal. {\it Science} $\mathbf{347}$, 294-298 (2015).
	

\bibitem{Huang2015}S.-M. Huang, S.-Y. Xu, I. Belopolski, C.-C. Lee, G. Chang, B. Wang, N. Alidoust,  G. Bian, M. Neupane, C. Zhang, S. Jia,  A. bansil. H. Lin, and M. Z. Hasan, A Weyl Fermion semimetal with surface Fermi arcs in the transition metal monopnictide TaAs class. \textit{Nat. Commun.} $\mathbf{6}$, 7373 (2015).

\bibitem{Weng2015} H. Weng, C. Fang, Z. Fang, A. Bernevig and X. Dai, Weyl Semimetal Phase in Noncentrosymmetric Transition-Metal Monophosphides. \textit{Phys. Rev. X} $\mathbf{5}$, 011029 (2015).

\bibitem{Hasan_TaAs}S.-Y. Xu, I. Belopolski, N. Alidoust, M. Neupane, G. Bian, C. Zhang, R. Sankar, G. Chang, Z. Yuan,  C.-C. Lee, S.-M. Huang, H. Zheng,J. Ma, D. S. Sanchez, B. Wang, A. Bansil, F. Chou, P. P. Shibayev, H. Lin, S. Jia, and M. Z. Hasan, Discovery of a Weyl fermion semimetal and topological Fermi arcs. \textit{ Science} $\mathbf{349}$, 613-617 (2015).

\bibitem{MIT_Weyl} L. Lu \textit{et al.}, Experimental observation of Weyl points \textit{ Science} $\mathbf{349}$, 622-624 (2015).

\bibitem{TaAs_Ding} B. Q. Lv  \textit{et al.}, Experimental Discovery of Weyl Semimetal TaAs. \textit{Phys. Rev. X} $\mathbf{5}$, 031013 (2015).
\bibitem{Chiral_anomaly_ChenGF} X. Huang  \textit{et al.}, Observation of the chiral anomaly induced negative magneto-resistance in 3D Weyl semi-metal TaAs. \textit{Phys. Rev. X} $\mathbf{5}$, 031023 (2015).
\bibitem{Chiral_anomaly_Jia} C. Zhang  \textit{et al.}, Observation of the Adler-Bell-Jackiw chiral anomaly in a Weyl semimetal. Preprint at http://arxiv.org/abs/1503.02630 (2015).
\bibitem{Semimetal_Jia} C. Zhang  \textit{et al.}, Tantalum Monoarsenide: an Exotic Compensated Semimetal. Preprint at http://arxiv.org/abs/1502.00251 (2015).


\bibitem{Weyl-SC-2} P. Hosur and  X.-L. Qi, Time-reversal invariant topological superconductivity in doped Weyl semimetals.\textit{Phys. Rev. B.} $\mathbf{90}$, 045130 (2014).
\bibitem{Weyl-SC-5} G. Bednik, A. A. Zyuzin and A. A. Burkov,  Superconductivity in Weyl semimetals. \textit{Phys. Rev. B.} $\mathbf{92}$, 035153 (2015).
\bibitem{Weyl-SC-7} Y. Li  and F. D. M. Haldane, Topological nodal Cooper pairing in doped Weyl seimetals. Preprint at http://arxiv.org/abs/1510.01730 (2015).



\bibitem{TaAs_Ding_2} B. Q. Lv \textit{et al.}, Observation of Weyl nodes in TaAs. \textit{Nat. Phys.} $\mathbf{11}$, 724-727 (2015).

\bibitem{TaAs_Chen}L. X. Yang  \textit{et al.}, Weyl semimetal phase in the non-centrosymmetric compound TaAs. \textit{Nat. Phys.} $\mathbf{11}$, 728-732 (2015).
\bibitem{NbAs_Hasan}S.-Y. Xu  \textit{et al.}, Discovery of a Weyl fermion state with Fermi arcs in niobium arsenide. \textit{Nature Phys.} $\mathbf{11}$, 748-755 (2015).

\bibitem{TaP_Shi} N. Xu  \textit{et al.}, Observation of Weyl nodes and Fermi arcs in TaP. Preprint at http://arxiv.org/abs/1507.03983 (2015).

\bibitem{TaP_Hasan} S.-Y. Xu  \textit{et al.}, Experimental discovery of a topological Weyl semimetal state in TaP. \textit{Science Advances} $\mathbf{1}$, e1501092 (2015).
\bibitem{NbP_DLFeng} D. F. Xu  \textit{et al.}, Observation of Fermi Arcs in non-Centrosymmetric Weyl Semi-metal Candidate NbP. Preprint at http://arxiv.org/abs/1509.03847 (2015).
\bibitem{NbP_Hasan} I. Belopolski, S.-Y. Xu, D. Sanchez, G. Chang, C. Guo, M. Neupane, H. Zheng, C.-C. Lee, S.-M. Huang, G. Bian, N. Alidoust, T.-R. Chang, B. Wang, X. Zhang, A. Bansil, H.-T. Jeng, H. Lin, S. Jia, and M. Z. Hasan, \textit{Phys. Rev. Lett.}, in press, Preprint at http://arxiv.org/abs/1509.07465 (2015).
\bibitem{NbP_Ando} S. Souma \textit{et al.}, Direct Observation of Nonequivalent Fermi-Arc States of Opposite Surfaces in Noncentrosymmetric Weyl Semimetal NbP. Preprint at http://arxiv.org/abs/1510.01503 (2015).
\bibitem{NbP_Chen}Z. K. Liu  \textit{et al.}, Evolution of the Fermi surface of Weyl semimetals in the transition metal pnictide family. \textit{Nat. Mater.} $\mathbf{15}$, 27-31 (2016).

\bibitem{NbP_STM} H. Zheng, S.-Y. Xu, G. Bian, C. Guo, G. Chang, D. S.Sanchez, I. Belopolski, C.-C. Lee, S.-M. Huang, X. Zhang, R. Sankar, N. Alidoust, T.-R. Chang, F. Wu, T. Neupert, F.Chou, H.-T. Jeng, N. Yao, A. Bansil, S. Jia, H. Lin, and M. Z. Hasan, \textit{ACS nano.} $\mathbf{10}$, 1378(2016).

\bibitem{MoWTe2} T.-R. Chang \textit{et al.}, Arc-tunable Weyl Fermion metallic state in Mo$_x$W$_{1-x}$Te$_2$. \textit{Nat. Commum.}, in press,  Preprint at http://arxiv.org/abs/1508.06723 (2015).

\bibitem{Ta3S2_Crystal_1} S. Kim, K. S. Nanjundaswamy and  T. Hughbanks, \textit{\textbf{Single-crystal}} structure of tantalum sulfide (Ta$_3$S$_2$). Structure and bonding in the Ta$_6$Sn $(n = 1,3,4,5)$ pentagonal-antiprismatic chain compounds. \textit{Inorg. Chem.} $\mathbf{30}$, 159-164 (1991).
\bibitem{Ta3S2_Crystal_2} H. Nozaki, W. Hiroaki  and  S. Takekawa, Properties of Ta$_3$S$_2$, Ta$_2$S and Ta$_6$S. \textit{J. Phys. Soc. Japan} $\mathbf{60}$, 3510-3515 (1991).


\bibitem{WTe2} A. A. Soluyanov \textit{et al.}, Type-II Weyl semimetals. \textit{Nature} $\mathbf{527}$, 495-498 (2015)

\bibitem{Zheng} H. Zheng \textit{et al.}, Discrete plasticity in sub-10-nm-sized gold crystals. \textit{Nat. Commun.} $\mathbf{1}$, 144 (2015)


\bibitem{Vanderbilt} J. Liu,  and D. Vanderbilt, Weyl semimetals from noncentrosymmetric topological insulators. \textit{Phys. Rev. B} $\mathbf{90}$, 155316 (2014).
\bibitem{BiX} C.-C. Liu,  J.-J. Zhou, Y. Yao and F. Zhang, Weak Topological Insulators and Composite Weyl Semimetals: $\beta$-Bi$_4$X$_4$ (X$=$Br, I). Preprint at http://arxiv.org/abs/1509.07183 (2015).

\bibitem{Ozaki_1} T.  Ozaki, Variationally optimized atomic orbitals for large-scale electronic structures. \textit{Phys. Rev. B} $\mathbf{67}$, 155108 (2003).
\bibitem{Ozaki_2} T.  Ozaki \textit{et al.}, OpenMX V.3.7 (2013) http://www.openmx-square.org.
\bibitem{Theurich} G. Theurich  and N. A. Hill, Self-consistent treatment of spin-orbit coupling in solids using relativistic fully separable $ab initio$ pseudopotentials. \textit{Phys. Rev. B} $\mathbf{64}$, 073106 (2001).
\bibitem{Perdew} J. P. Perdew, K. Burke, and M. Ernzerhof, Generalized Gradient Approximation Made Simple. \textit{Phys. Rev. Lett.} $\mathbf{77}$, 3865-3868 (1996).







%
%

\end{thebibliography}
\end{document}